\documentstyle[amstex,amsfonts,aps,prb,graphicx,epsf]{revtex}

\begin{document}
\draft
\title{X-boson cumulant approach to the periodic Anderson model}
\author{R. Franco, M. S. Figueira}
\address{Instituto de F\'{\i}sica, C.P. 100.093\\
Universidade Federal Fluminense,UFF\\
Av. Litor\^{a}nea s/$n^{\underline{o}}$, 24210-340 Niter\'{o}i, Rio de
Janeiro, Brasil}
\author{M. E. Foglio}
\address{Instituto de F\'{\i}sica ``Gleb Wataghin''\\
Universidade Estadual de Campinas,UNICAMP\\
13083-970 Campinas, S\~{a}o Paulo, Brasil}
\date{\today }
\maketitle

\begin{abstract}
The Periodic Anderson Model (PAM) can be studied in the limit $U=\infty $ by
employing the Hubbard $X$ operators to project out the unwanted states. We
have already studied this problem employing the cumulant expansion with the
hybridization as perturbation, but the probability conservation of the local
states (completeness) is not usually satisfied when partial expansions like
the ``Chain Approximation (CHA)'' are employed. Here we treat the problem by
a technique inspired in the mean field approximation of Coleman's
slave-bosons method, and we obtain a description that avoids the unwanted
phase transition that appears in the mean-field slave-boson method both when 
$\mu >>E_{f}$ at low $T$ and for all parameters at intermediate temperatures.
\end{abstract}


\begin{description}
\item[Short Title]  X-boson cumulant approach to the periodic Anderson model
\end{description}

\twocolumn

\narrowtext

\section{INTRODUCTION}

{\ \label{Sec1} }

{\ In the limit of infinite Coulomb repulsion ($U=\infty $ ) one can
transform the Periodic Anderson Model (PAM) with two channel ($SU(2)$) by
employing the Hubbard $X_{j,ab}$=$\left| j,a\right\rangle \left\langle
j,b\right| $ operators,\cite{FFM} where the set $\left\{ \left|
j,a\right\rangle \right\} $ is an orthonormal basis in the space of
interest. Projecting out the components with more than one electron from any
local state at site $j$ one obtains 
\begin{eqnarray}
H &=&\sum_{{\bf k},\sigma }E_{{\bf k},\sigma }c_{{\bf k},\sigma }^{\dagger
}c_{{\bf k},\sigma }+\sum_{j,\sigma }\ E_{f,j\sigma }X_{j,\sigma \sigma } 
\nonumber \\
&&+\sum_{j,\sigma ,{\bf k}}\left( V_{j,{\bf k},\sigma }X_{j,0\sigma
}^{\dagger }c_{{\bf k},\sigma }+V_{j,{\bf k},\sigma }^{\ast }c_{{\bf k}%
,\sigma }^{\dagger }X_{j,0\sigma }\right) .  \label{Eq.3}
\end{eqnarray}
}

{\ \noindent\ The first term is the Hamiltonian of the conduction electrons (%
$c$-electrons). The second term describes independent localized electrons ($%
f $-electrons), and a simple index $j$ is used to indicate the sites. The
last term is the hybridization Hamiltonian giving the interaction between
the $c$-electrons and the $f$-electrons with $V_{j,{\bf k},\sigma }=(1/\sqrt{%
N_{s}})V_{\sigma }({\bf k})\exp {(i{\bf k}.{\bf R}_{j})}$, where ${{\bf R}%
_{j}}$ is the position of site $j$ and $N_{s}$ the number of sites. Notice
that this interaction conserves the spin component $\sigma $. }

{\ The $X$ operators do not satisfy the usual commutation relations. One has
then to use the product rules: } {\ 
\begin{equation}
X_{j,ab}.X_{j,cd}=\delta _{b,c}X_{j,ad},  \label{HO}
\end{equation}
and the diagrammatic methods based on Wick's theorem are therefore not
applicable. We shall use instead a cumulant expansion that was originally
employed by Hubbard\cite{Hubard4} to study his model, and that was latter
extended to the PAM.\cite{FFM,Infinite} }

{\ There are four local states at each site $j$ of the lattice: the vacuum
state \ $\left| \ j,0\right\rangle $, the two states $\left| \ j,\sigma
\right\rangle $ of one electron with spin component $\sigma $, and the state 
$\left| \ j,2\right\rangle $ with two local electrons.\ When $U\rightarrow
\infty $ \ the state $\left| \ j,2\right\rangle $ is empty, and we have used
the Hubbard operators to project it out from the space of local states at
site $j$. In this space, the identity $I_{j}$ at site $j$ should then
satisfy the relation: 
\vspace{0.5cm}
\begin{equation}
X_{j,00}+X_{j,\sigma \sigma }+X_{j,\overline{\sigma }\overline{\sigma }%
}=I_{j},  \label{Eq.1}
\end{equation}
where $\overline{\sigma }$ is the spin component opposite to $\sigma ,$ and
the three $X_{j,aa}$ are the projectors into $\mid j,a\rangle $. The
occupation numbers $n_{j,a}=<X_{j,aa}>$ can be calculated from appropriate
Green's functions (GF), and assuming translational invariance we can write $%
n_{j,a}=n_{a}$ (independent of j), so that from Eq. (\ref{Eq.1}) it follows
that 
\begin{equation}
n_{0}+n_{\sigma }+n_{\overline{\sigma }}=1.  \label{Eq.2}
\end{equation}
\vspace{0.5cm}
We shall call this relation ``completeness'', and it has been found that it
is not usually satisfied when the $n_{a}$ are calculated with approximate
cumulant Green's functions (GF).\cite{Ufinito} An approximation displaying
this behavior is the ``Chain Approximation'' (CHA), which was first employed
by Hewson,\cite{Hewson,Enrique} and is the most general cumulant expansion
with only second order cumulants, as well as being $\Phi -$derivable. \cite
{Baym61,ChainPhi} In a previous work\cite{ChainPhi} we have stated a
conjecture that gives a systematic way of achieving completeness by adding a
set of diagrams to an arbitrary family. This result was verified for the
CHA, and the resulting set of diagrams is not $\Phi -$derivable any more. }

{\ In the present work we recover completeness by a procedure similar to
that employed by Coleman \cite{Coleman84} in the slave boson approach. This
last method predicts unphysical second order phase transitions both for all
parameters at intermediate temperatures }$T${\ and in the Kondo region $(\mu
>>E_{f,j\sigma }=E_{f})$ at low }$T${. Coleman\cite{Coleman87} has observed
that these effects are artifacts of the theory, and the advantage of the
present treatment is that those spurious phase transitions do not occur. Our
method gives results very close to those obtained by the slave boson in the
Kondo limit at low temperatures, while it recovers those of the CHA at high }%
$T${\ for all parameters. }

\section{X-boson Cumulant Method}

{\ \label{Sec2} }

{\ The present work is a modification of a preliminary version \cite
{JMMM2001} that was inspired in the mean-field approximation of Coleman's
``slave boson'' method.\cite{Coleman84,Read87} He writes the Hubbard X
operators as a product of ordinary bosons and fermions: $X_{j,oo}\rightarrow
b_{j}^{+}b_{j}$, $X_{j,o\sigma }\rightarrow b_{j}^{+}f_{j,\sigma }$, $%
X_{j,\sigma o}\rightarrow f_{j,\sigma }^{+}b_{j}$, and uses the equivalent
of our Eq. (\ref{Eq.2}) to avoid states with more than one electron at each
site $j$. In the spirit of the mean field approximation $b_{i}^{+}%
\rightarrow <b_{i}^{+}>=r$, and the method of Lagrangian multipliers is then
used to find the ``best'' Hamiltonian that satisfies Eq.~(\ref{Eq.2}). The
problem is then reduced to an uncorrelated Anderson lattice with
renormalized hybridization $V\rightarrow rV$ and $f$ level $\epsilon
_{f}\rightarrow \epsilon _{f}+\lambda $. }

{\ Our method consists in adding the product of each Eq. (\ref{Eq.1}) times
a Lagrange multiplier $\Lambda _{j}$ to Eq.~(\ref{Eq.3}), and the new
Hamiltonian generates the functional that we shall minimize employing
Lagrange's method. Instead of the parameter $r$ we introduce 

\vspace{0.5cm}
\begin{equation}
R\equiv <X_{j,oo}>=<b_{j}^{+}b_{j}>,  \label{Eq.5}
\end{equation}

and we call the method ``X-boson'' because the Hubbard operator $X_{j,oo}$
has a ``Bose-like'' character.\cite{FFM} }
{\ Considering Eqs.~(\ref{Eq.1},\ref{Eq.5}), and employing a constant
hybridization $V$, as well as site independent local energies ${E}%
_{f,j,\sigma }={E}_{f,\sigma }$ and Lagrange parameters $\Lambda
_{j}=\Lambda $, we obtain a new Hamiltonian with the same form of Eq.~(\ref
{Eq.3}) 
\begin{eqnarray}
H &=&\sum_{{\bf k},\sigma }\ E_{{\bf k},\sigma }\ c_{{\bf k},\sigma
}^{\dagger }c_{{\bf k},\sigma }  \nonumber \\
&&+\sum_{j,\sigma }\tilde{E}_{f,\sigma }X_{j,\sigma \sigma }+N_{s}\Lambda
(R-1)  \nonumber \\
&&+V\sum_{j,{\bf k},\sigma }\left( X_{j,0\sigma }^{\dagger }\ c_{{\bf k}%
,\sigma }+c_{{\bf k},\sigma }^{\dagger }\ X_{j,0\sigma }\right) ,
\label{Eq.7}
\end{eqnarray}
\noindent but with a renormalized localized energy 
\begin{equation}
\tilde{E}_{f,\sigma }=E_{f,\sigma }+\Lambda .  \label{Energ}
\end{equation}
This procedure allows for an independent variation of $R$ when the Free
Energy is minimized, even though the completeness relation 
\begin{equation}
R=1-\sum_{\sigma }<X_{\sigma \sigma }>  \label{Eq. 9}
\end{equation}
must be simultaneously satisfied. }

{\ In the slave-boson approach a one-body Hamiltonian is obtained at this
stage, and Eq.~(\ref{Eq.2}) is then necessary to avoid the occupation of
more than one localized electron per site, while the conservation of
probability in the space of the local states is automatically satisfied,
because normal fermion operators are employed in the transformed
Hamiltonian. The Eq.~(\ref{Eq.7}) on the other hand employs $X$-operators,
that force the local states to be singly occupied at most, but Eq.~(\ref
{Eq.2}) (completeness) must be imposed here because it is not automatically
satisfied when approximate GF are employed to calculate the $n_{a}$. In the
present work we shall employ the GF of the Chain approximation (CHA),\cite
{Hewson,Enrique} because they give a fair description of the system in spite
of their simplicity. }

{\ The present treatment employs the Grand Canonical Ensemble of electrons,
and instead of Eq.~(\ref{Eq.3}) we shall use 
\begin{equation}
{\cal H}=H-\mu \left\{ \sum_{{\bf k,\sigma }}c_{{\bf k,\sigma }}^{\dagger
}c_{{\bf k,\sigma }}+\sum_{ja}\nu _{a}X_{j,aa}\right\} \qquad ,  \label{E2.2}
\end{equation}
}

{\ \noindent where $\nu _{a}=0,1$ is the number of electrons in state $\mid
a>$. It is then convenient to define 
\begin{equation}
\varepsilon _{j,a}=E_{f,j,a}-\mu \nu _{a}  \label{E2.3a}
\end{equation}
and 
\begin{equation}
\varepsilon _{{\bf k\sigma }}=E_{{\bf k\sigma }}-\mu \ ,  \label{E2.3b}
\end{equation}
because }$E_{f,j,a}${\ and $E_{{\bf k,\sigma }}$ appear only in that form in
all the calculations. The exact and unperturbed averages of the operator $A$
are denoted in what follows by $<A>_{{\cal H}}$ and \mbox{$<A>$}
respectively. }

\section{The Chain Approximation Green's Functions}

{\ \label{Sec3} The CHA gives simple but useful approximate propagators,
obtained in the cumulant expansion by taking the infinite sum of all the
diagrams that contain ionic vertices with only two lines. The laborious
calculation of the general treatment\cite{FFM} is rather simplified in this
case, and we shall give a brief description of the technique, particularly
when only the imaginary time is Fourier transformed, because this
intermediate situation is not discussed in reference \onlinecite
{FFM} and it is necessary to calculate the impurity problem. The only $%
X_{\alpha }$ and $X_{\alpha }^{\dagger }$ operators of the Fermi type that
appear in the calculation have $\alpha =(0,\sigma )$, and the f-electron GF
in real space and imaginary time are 
\begin{equation}
G^{ff}(j,\alpha ,\tau ;j\prime ,\alpha \prime ,\tau \prime )=\left\langle
\left( X_{j,\alpha }(\tau )X_{j\prime ,\alpha \prime }^{\dagger }(\tau
\prime )\right) _{+}\right\rangle _{{\cal H}},  \label{Eqn.1}
\end{equation}
where $\widehat{X}_{j,\alpha }(\tau )=\exp \left( \tau {\cal H}\right)
X_{j,\alpha }\exp \left( -\tau {\cal H}\right) $ corresponds to the
Heisenberg representation and the subindex $+$ indicates that the operators
inside the parenthesis are taken in the order of increasing $\tau $ to the
left, with a change of sign when the two Fermi-type operators have to be
exchanged to obtain this ordering.\cite{FFM} \ In a similar way one defines
the c-electron GF $G^{cc}({\bf k},\sigma ,\tau ;{\bf k\prime },\sigma \prime
,\tau \prime )$ as well as the mixed GF $G^{fc}(j,\alpha ,\tau ;{\bf k\prime 
},\sigma \prime ,\tau \prime )$ and $G^{cf}({\bf k},\sigma ,\tau ;j\prime
,\alpha \prime ,\tau \prime )$. }

{\ The boundary conditions of these GF with respect to $\tau $ makes
possible to expand them in Fourier series\cite{FFM} employing the Matsubara
frequencies $\omega _{n}={(2n+1)i\pi }/{\beta }$, where $n$ is any integer.
Because of the invariance of ${\cal H}$ against $\tau $ translations we have
frequency conservation,\cite{FFM} so that $G^{ff}(j,\alpha ,i\omega
_{n};j\prime ,\alpha \prime ,i\omega _{n}\prime )=0$ unless $\omega
_{n}+\omega _{n}\prime =0$, and one can then write 
\begin{equation}
G^{ff}(j,\alpha ,i\omega _{n};j\prime ,\alpha \prime ,i\omega _{n}\prime
)=G_{j\alpha ;j\prime \alpha \prime }^{ff}(z_{n})\ \Delta (\omega
_{n}+\omega _{n}\prime )  \label{Eqn.2}
\end{equation}
and similar relations for the $G^{cc}$,$G^{fc}$,} {and $G^{cf}$. Here we
have used }$z_{n}\equiv i\omega _{n}${, and we shall keep this notation in
what follows, as well as employing $\Delta (\gamma )\equiv \delta _{\gamma
,0}$ (a modified notation for Kr\"{o}necker's delta). The ``bare GF'' (i.e.
with all $V_{j,{\bf k},\sigma }=0$) take a fairly simple form \ because many
of the relevant operators become statistically independent, and we then have 
\begin{eqnarray}
&&G^{ff}(j,\alpha ,z_{n};j\prime ,\alpha \prime ,z_{n}\prime )\Longrightarrow
\nonumber \\
&&\hspace{1cm}G_{f,\alpha }^{0}(z_{n})\ \Delta (\omega _{n}+\omega
_{n}\prime )\ \delta _{j,j\prime }\ \delta _{\alpha ,\alpha \prime },
\label{Eqn.4a} \\
&&G^{cc}({\bf k},\alpha ,z_{n};{\bf k}\prime ,\alpha \prime ,z_{n}\prime
)\Longrightarrow  \nonumber \\
&&\hspace{1cm}G_{c,{\bf k}\sigma }^{0}(z_{n})\ \Delta (\omega _{n}+\omega
_{n}\prime )\ \delta _{{\bf k},{\bf k}\prime }\ \delta _{\sigma ,\sigma
\prime },  \label{Eqn.4b} \\
&&G^{fc}\Longrightarrow G^{cf}\Longrightarrow 0.  \label{Eqn.4c}
\end{eqnarray}
The$\ \delta _{\alpha ,\alpha \prime }$ in the bare GF follows from the
commutation of ${\cal H}$ with the $z$ component of the spin. }

\begin{figure}[ht]
\centering
\epsfysize=2.3truein \centerline  {\epsfbox{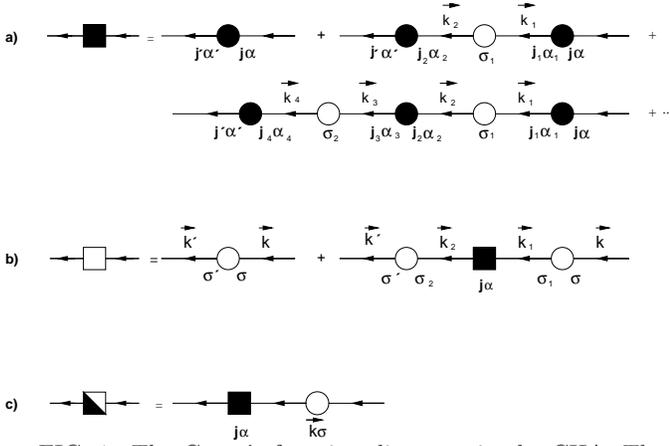}}
\caption{The Green's function diagrams in the CHA. The filled
circles (vertices) corresponds to the $f$-electron cumulants and the empty
ones to those of the $c$-electrons. The lines (edges) joining two vertices
represent the perturbation (hybridization) (cf. Section~\ref{Sec3}) a)
Diagrams for the $f$-electron GF in the CHA. The CHA diagram for the $f$%
-electron $G_{j_{i}\protect\alpha _{i},j_{f}\protect\alpha _{f}}^{ff}(i%
\protect\omega _{n})$ is represented by the filled square to the left. b)
Diagrams for the $c$ electron GF in the CHA. The $G_{{\bf k\prime ,}\protect%
\sigma \prime ;{\bf k,}\protect\sigma }^{cc}(z_{n})$ is represented by the
square symbol to the left. c) Diagrams for the $f$-$c$ electron GF in the
CHA. The $G_{{\bf j};{\bf k,}\protect\sigma }^{fc}(z_{n})$ is represented by
the square symbol to the left.}
\label{GffCHA}
\end{figure}

{\ The diagrams in real space that contribute to the CHA are schematized in
Fig~(\ref{GffCHA}). The meaning of the symbols in the cumulant diagrams is
the following:\newline
a) the ``vertex'' %\newline
\mbox{\put(9,3){\circle*{10.7}}\hspace{16pt} = $G_{f,0\sigma
}^{o}(z_{n})=-D_{0\sigma }/{(z_{n}-\varepsilon _{f,\sigma
})}$}\newline
is the $f$ bare cumulant GF , where \newline
$D_{0\sigma }=<X_{oo}>+<X_{\sigma \sigma }>$} \newline
{b) the ``vertex'' 
\mbox{\put(9,3){\circle{10.7}} \hspace{16pt}
$=G_{c,{\bf k}\sigma }^{o}(z_{n})=-1/{(z_{n}-\varepsilon _{{\bf
k\sigma }})}$}\newline
is the $c$ bare cumulant GF \newline
c) the lines (edges)$\ $ determine an open loop with a definite direction.
When the line points to the f-vertex is $\longleftarrow $=$V_{j,{\bf k}%
,\sigma }$, while $\longleftarrow $=$V_{j,{\bf k},\sigma }^{\ast }$ when it
points to the conduction vertex \newline
d) The cumbersome sign and symmetry factors\cite{FFM,Hubard4} are rather
simple in the CHA: it is only necessary to multiply the $G^{fc}$ and the $%
G^{cf}$ into a minus sign. \newline
e) As both the $\omega $ and the $\sigma $ are conserved along the open
loop, it only remains to sum over all the internal $j$ and ${\bf k}$. }

{\ We then obtain (with $\alpha =(0\sigma )$): 
\begin{eqnarray}
&&G_{j\prime \alpha \prime ;j\alpha }^{ff}(z_{n})=\delta _{\alpha \prime
,\alpha }G_{f,\alpha }^{o}(z_{n})  \nonumber \\
&&+G_{f,\alpha }^{o}(z_{n})\sum_{{\bf k}_{1}}V_{j\prime ,{\bf k}_{1}\sigma
}G_{c,{\bf k}_{1}\sigma }^{o}(z_{n})V_{j,{\bf k}_{1}\sigma }^{\ast
}G_{f,\alpha }^{o}(z_{n})  \nonumber \\
&&+G_{f,\alpha }^{o}(z_{n})\sum_{{\bf k}_{2}}V_{j\prime ,{\bf k}_{2}\sigma
}G_{c,{\bf k}_{2}\sigma }^{o}(z_{n})\sum_{j_{1}}V_{j_{1},{\bf k}_{2}\sigma
}^{\ast }G_{f,\alpha }^{o}(z_{n})  \nonumber \\
&&\times \sum_{{\bf k}_{1}}V_{j_{1},{\bf k}_{1}\sigma }G_{c,{\bf k}%
_{1}\sigma }^{o}(z_{n})V_{j,{\bf k}_{1}\sigma }^{\ast }G_{f,\alpha
}^{o}(z_{n})+....\qquad .  \label{Eqn.5}
\end{eqnarray}
}

\subsubsection{The impurity case}

{\ \bigskip When there is a single impurity at a given $j$, each of the sums
over the internal $j_{r}$ reduces to a single term with $j\prime =j$, and we
introduce 
\begin{eqnarray}
M_{\sigma }(z_{n}) &=&\sum_{{\bf k}_{1}}V_{j{\bf , k}_{1}\sigma }G_{c,{\bf k}%
_{1}\sigma }^{o}(z_{n})V_{j,{\bf k}_{1}\sigma }^{\ast }  \nonumber \\
&=&\frac{1}{N_{s}}\sum_{{\bf k}}\left| V_{\sigma }({\bf k})\right| ^{2}G_{c,%
{\bf k}\sigma }^{o}(z_{n}),  \label{Eqn.6}
\end{eqnarray}
which is the local GF at site $j$ times $\left| V_{\sigma }\right| ^{2}$
when\ the hybridization is purely local, i.e. when $V_{\sigma }({\bf k}%
)=V_{\sigma }$. The Eq. (\ref{Eqn.5}) is then a geometric series that is
easily summed: 
\begin{eqnarray}
&&G_{j\prime \alpha \prime ;j\alpha }^{ff}(z_{n})=\delta _{\alpha \prime
,\alpha }\delta _{j\prime ,j}\frac{G_{f,\alpha }^{o}(z_{n})}{1-G_{f,\alpha
}^{o}(z_{n})\ M_{\sigma }(z_{n})}  \label{Eqn.7} \\
&&\qquad =\delta _{\alpha \prime ,\alpha }\delta _{j\prime ,j}\frac{%
-D_{0\sigma }}{z_{n}-{\varepsilon _{f,\sigma }}+D_{0\sigma }\ M_{\sigma
}(z_{n})}.  \label{Eqn.7a}
\end{eqnarray}
}

{\ Employing the same technique, we obtain for the CHA the $G^{cc}$, that
gives the scattering by the local impurity of a conduction electron\ ${\bf k,%
}\sigma $ into ${\bf k\prime ,}\sigma \prime $:\bigskip 
\begin{eqnarray}
&&G_{{\bf k\prime}\sigma \prime ;{\bf k}\sigma }^{cc}(z_{n})=\delta
_{\sigma \prime ,\sigma }\Bigl\{G_{c,{\bf k}\sigma }^{o}(z_{n})\ \delta _{%
{\bf k},{\bf k}\prime }  \nonumber \\
&&\left. +G_{c,{\bf k\prime }\sigma }^{o}(z_{n})V_{\sigma }^{\ast }({\bf %
k\prime })G_{j\alpha ;j\alpha }^{ff}(z_{n})\times V_{\sigma }({\bf k})G_{c,%
{\bf k}\sigma }^{o}(z_{n})\right\} ,  \label{Eqn.8}
\end{eqnarray}
as well as the mixed GF (with $\alpha ^{\prime }=(0\sigma ^{\prime })$) 
\begin{equation}
G_{j{\bf \prime \alpha \prime };{\bf k}\sigma }^{fc}(z_{n})=\delta _{\sigma
\prime ,\sigma }\delta _{j\prime ,j}G_{j;{\bf k}\sigma }^{fc}(z_{n}),
\label{Eqn.9}
\end{equation}
where we have already included the minus sign (discussed in rule d) above)
into }

\begin{eqnarray}
&&G_{j;{\bf k}\sigma }^{fc}(z_{n})=-G_{j\alpha ;j\alpha }^{ff}(z_{n})V_{j,%
{\bf k}\sigma }G_{c,{\bf k}\sigma }^{o}(z_{n})  \label{Eqn.9a} \\
&=&-\frac{D_{0\sigma }V_{j,{\bf k}\sigma }}{z_{n}-{\varepsilon _{f,\sigma }+%
}D_{0\sigma }M_{\sigma }(z_{n})}\ \times \frac{1}{z_{n}-{\varepsilon _{{\bf %
k\sigma }}}}.  \label{Eqn.9b}
\end{eqnarray}

\subsubsection{The lattice case}

{\ The case of the GF in reciprocal space and imaginary frequencies has been
discussed in detail in reference,\cite{FFM} and in the CHA one follows the
same prescriptions given above, but replacing the sum over internal $j$ and $%
{\bf k}$ by a conservation of ${\bf k}$ along the whole chain, so that we
have 
\begin{eqnarray}
&&G^{ff}({\bf k}\prime ,(0\sigma \prime ),z_{n}\prime ;{\bf k},(0\sigma
),z_{n})  \nonumber \\
&&\hspace{1cm}=\delta _{{\bf k},{\bf k}\prime }\ \delta _{\sigma ,\sigma
\prime }\ \Delta (\omega _{n}+\omega _{n}\prime )G_{{\bf k}{\sigma }%
}^{ff}(z_{n}),  \label{Eqn.10a} \\
&&G^{cc}({\bf k}\prime ,\sigma \prime ,z_{n}\prime ;{\bf k},\sigma ,z_{n}) 
\nonumber \\
&&\hspace{1cm}=\delta _{{\bf k},{\bf k}\prime }\ \delta _{\sigma ,\sigma
\prime }\ \Delta (\omega _{n}+\omega _{n}\prime )G_{{\bf k}{\sigma }%
}^{cc}(z_{n}),  \label{Eqn.10b} \\
&&G^{fc}({\bf k}\prime ,(0\sigma \prime ),z_{n}\prime ;{\bf k},\sigma ,z_{n})
\nonumber \\
&&\hspace{1cm}=\delta _{{\bf k},{\bf k}\prime }\ \delta _{\sigma ,\sigma
\prime }\ \Delta (\omega _{n}+\omega _{n}\prime )G_{{\bf k}{\sigma }%
}^{fc}(z_{n}),  \label{Eqn.10c}
\end{eqnarray}
where 
\begin{equation}
G_{{\bf k}{\sigma }}^{ff}(z_{n})=\frac{-D_{\sigma }\left( z_{n}-\varepsilon
_{{\bf k}\sigma }\right) }{\left( z_{n}-\varepsilon _{f,\sigma }\right)
\left( z_{n}-\varepsilon _{{\bf k}\sigma }\right) -|V_{\sigma }({\bf k}%
)|^{2}D_{\sigma }},  \label{Eqn.11}
\end{equation}
\begin{equation}
G_{{\bf k}{\sigma }}^{cc}(z_{n})=\frac{-\left( z_{n}-\varepsilon _{f,\sigma
}\right) }{\left( z_{n}-\varepsilon _{f,\sigma }\right) \left(
z_{n}-\varepsilon _{{\bf k}\sigma }\right) -|V_{\sigma }({\bf k}%
)|^{2}D_{\sigma }},  \label{Eqn.12}
\end{equation}
and 
\begin{equation}
G_{{\bf k}{\sigma }}^{fc}(z_{n})=\frac{-\ D_{\sigma }V_{\sigma }({\bf k})}{%
\left( z_{n}-\varepsilon _{f,\sigma }\right) \left( z_{n}-\varepsilon _{{\bf %
k}\sigma }\right) -|V_{\sigma }({\bf k})|^{2}D_{\sigma }}.  \label{Eqn.13}
\end{equation}
}

\subsubsection{The Slave Boson GF vs the GF of the CHA}

{\label{Sec3.3} }

The Slave Boson GF in the mean field approximation (cf. references %
\onlinecite{Read87,FoglioSUN,Hewson93}) are just the GF of the uncorrelated
problem ($U=0$) but with a renormalized hybridization {$V\rightarrow 
\overline{V}\equiv rV$ plus a condition that forces the local electron to an
occupation less or equal to one. The GF of the CHA given above are formally
very close to the uncorrelated ones, but they can not be reduced to them by
any change of scale (except for ${D}_{\sigma }=1$, when we recover the
slave-boson Green's functions if we put $V\rightarrow rV=\overline{V}$ in
Eqs.~(\ref{Eqn.7}-\ref{Eqn.9b}) and Eqs.~(\ref{Eqn.10a}-\ref{Eqn.13})). The
obvious change }$\sqrt{D_{0\sigma }}V\rightarrow \overline{V}$, leaves an
extra factor $D_{0\sigma }$ in the $G_{j\prime \alpha \prime ;j\alpha
}^{ff}(z_{n})$ and $G_{{\bf k}{\sigma }}^{ff}(z_{n})$, as well as a $\sqrt{%
D_{0\sigma }}$ in both $G_{{\bf j};{\bf k,}\sigma }^{fc}(z_{n})$\ and $G_{%
{\bf k}{\sigma }}^{fc}(z_{n})$, and these factors are responsible for the
correlation in this approximation, and lead to essential differences with
the uncorrelated case.\cite{FoglioSUN} In particular, they force the total
occupation $n_{f}$ of the f electron to $n_{f}\leq 1$, while in the
uncorrelated case the relation $n_{f}\leq 2$ is satisfied. In the slave
boson method the imposed condition $n_{f}\leq 1$ is fulfilled by a shift in
the local energy $\varepsilon _{f,\sigma }\rightarrow \widetilde{\varepsilon 
}_{f,\sigma }\equiv \varepsilon _{f,\sigma }+\lambda $ and a reduction of
the hybridization to {$V\rightarrow \widetilde{{V}}\equiv rV$}. From an
operational point of view, a shift in $\varepsilon _{f,\sigma }$ might not
be sufficient to force $n_{f}\leq 1$ because the hybridization extends the f
spectral density to the whole conduction band, and reducing $V$ helps to
satisfy this condition. By increasing the temperature $T$ or the chemical
potential $\mu $ the value $\widetilde{V}=0$ is presently reached, leading
to a decoupling of the two type of electrons that can be interpreted, from a
more general point of view, as a change of phase related to a symmetry
breaking of the mean field Hamiltonian. Although the condition that forces
completeness in the CHA is identical to that employed in the slave boson
method to force $n_{f}\leq 1$, it has a rather different origin, being only
a consequence of using a reduced set of diagrams in the perturbative
expansion,\cite{ChainPhi} and the departures from completeness are usually
very moderate. In the formalism described in the present work, it is this
essential difference between the two methods that eliminates the spurious
phase transitions appearing in the slave boson method.

\section{The single impurity problem}

{\ \label{Sec4} }

{\ In the X-boson approach ${D}_{\sigma }=R+n_{f\sigma }$ must be calculated
self-consistently through minimization with respect to the parameter $R$ of
an adequate thermodynamic potential. When the total number of electrons }${N}%
_{t}$, the temperature $T$ and the volume $V_{s}$ are kept constant one
should minimize the Helmholtz free energy, but the same minimum is obtained
by employing the thermodynamic potential {$\Omega =-k_{B}T\ln ({\cal Q})$,
(where }${\cal Q}${\ is the Grand Partition Function) and keeping }$T$, $%
V_{s}$, and the chemical potential $\mu ${\ constant (this result is easily
obtained by} employing standard thermodynamic techniques).

A convenient way of calculating $\Omega $ is to employ the{\ }${\xi }${\
parameter integration method.\cite{Physica A 94}}$^{,}${\cite{Abrikosov}}$%
^{,}${\cite{Doniach} This method introduces a }$\xi $ dependent Hamiltonian $%
H({\xi })=H_{o}+{\xi }H_{1}${\ \ through a coupling constant }${\xi }${\
(with $0\leq {\xi }\leq 1$), where $H_{1}$ is the hybridization in our case.
For each }${\xi }$ there is an associated{\ thermodynamic potential $\Omega (%
{\xi })$ which satisfies: \ 
\begin{equation}
\left( \frac{\partial \Omega }{\partial {\xi }}\right) _{V_{s},T,\mu
}=<H_{1}({\xi })>_{{\xi }},  \label{Eqn.14}
\end{equation}
\ \noindent where $<A>_{{\xi }}$ is the ensemble average of an operator }$A$
for a{\ system with Hamiltonian $H({\xi })$ and the given values of $\mu ,$ $%
T$, and }$V_{s}${. Integrating this equation gives}

{\ 
\begin{equation}
\Omega =\Omega _{o}+\int_{0}^{1}d{\xi }<H_{1}({\xi })>_{{\xi }},
\label{Eqn.15}
\end{equation}
\noindent where }$\Omega _{o}$ is the thermodynamic potential of the system
with ${\xi }=0$. This value of ${\xi }$ corresponds to a system without
hybridization, and one obtains (in the absence of magnetic field $%
\varepsilon _{{\bf k\sigma }}=\varepsilon _{{\bf k}}$ and $\widetilde{%
\varepsilon }_{f\sigma }=\widetilde{\varepsilon }_{f}$) {\ 
\begin{eqnarray}
\Omega _{o} &=&-\frac{2}{\beta }{\sum_{{\bf k}}}\ln \left[ 1+\exp (-\beta
\varepsilon _{{\bf k}})\right]  \nonumber \\
&&-\frac{1}{\beta }\ln \left[ 1+2\exp (-\beta \widetilde{\varepsilon }_{f})%
\right] +\Lambda (R-1),  \label{Eqn.16}
\end{eqnarray}
} and to calculate $\Omega $ in Eq.(\ref{Eqn.15}) we use{\ 
\begin{equation}
<H_{1}>_{{\xi }}=2Re\left[ \sum_{{\bf k}\sigma }V_{j,{\bf k},\sigma }^{\ast
}\ \left\langle c_{{\bf k}\sigma }^{\dagger }X_{0\sigma }\right\rangle _{{%
\xi }}\right] .  \label{Eqn.17}
\end{equation}
} {The average }$\left\langle c_{{\bf k}\sigma }^{\dagger }X_{0\sigma
}\right\rangle _{{\xi }}$ {is obtained from the analytical continuation of
the Matsubara }$G_{{\bf j};{\bf k,}\sigma }^{fc}(z_{n},{\xi })\rightarrow $ $%
\overline{G}_{{\bf j};{\bf k,}\sigma }^{fc}(z,{\xi })$\ {\ }into the complex
upper and lower semiplanes, where $G_{{\bf j};{\bf k,}\sigma }^{fc}(z_{n},{%
\xi })$ is the GF in Eq.~(\ref{Eqn.9a}) but for $V_{j,{\bf k},\sigma }${$%
\rightarrow $}${\xi }V_{j,{\bf k},\sigma }${. One then finds 
\begin{align}
& \left\langle c_{{\bf k},\sigma }^{\dagger }X_{j,0\sigma }\right\rangle _{{%
\xi }}=\frac{i}{2\pi }\int\limits_{-\infty }^{\infty }d\omega \ \
n_{F}(\omega )  \nonumber \\
& \times \left\{ \overline{G}_{j,{\bf k\sigma }}^{fc}(\omega +i0;{\xi })-%
\overline{G}_{j,{\bf k\sigma }}^{fc}(\omega -i0;{\xi })\right\} ,
\label{Eqn.18}
\end{align}
where }$n_{F}(x)=1/\left[ 1+\exp (\beta x)\right] ${\ is the Fermi-Dirac
distribution. From Eqs.~(\ref{Eqn.6},\ref{Eqn.17},\ref{Eqn.18}) we then
obtain} {\ 
\begin{align}
\left\langle H_{1}\right\rangle _{{\xi }}=& -\frac{1}{\pi }%
\int\limits_{-\infty }^{\infty }d\omega \ n_{F}(\omega )\   \nonumber \\
& \times \sum_{\sigma }\mathop{\rm Im}\left\{ \frac{{\xi }D_{0\sigma
}M_{\sigma }(\omega ^{+})}{\omega ^{+}-{\widetilde{\varepsilon} _{f}+}%
D_{0\sigma }{\xi}^{2}M_{\sigma }(\omega ^{+})}\right\} ,  \label{Eqn.18a}
\end{align}
where }$\omega ^{+}=\omega +i0$. {This equation has} an interesting scaling
property: it is equal to the corresponding expression of the uncorrelated
system for the scaled parameter $\overline{V}_{j,{\bf k},\sigma }=\sqrt{%
D_{0\sigma }}V_{j,{\bf k},\sigma }$ (it is enough to remember that by
replacing $D_{0\sigma }=1$ in the GF of the CHA one obtains the
corresponding GF of the uncorrelated system). Rather than performing the ${%
\xi }$ and $\omega $\ integrations, we shall use the value of the $\Omega
^{u}$ for the uncorrelated system with $\overline{V}_{j,{\bf k},\sigma }=%
\sqrt{D_{0\sigma }}V_{j,{\bf k},\sigma }$ and employ Eq. (\ref{Eqn.15}) to
calculate $\int_{0}^{1}d{\xi }<H_{1}^{u}({\xi })>_{{\xi }}=\Omega
^{u}-\Omega _{o}^{u}$, where 
\begin{eqnarray}
&&\Omega _{o}^{u}=\frac{-2}{\beta }\left[ {\sum_{{\bf k}}}\ln \left[ 1+\exp
(-\beta \varepsilon _{{\bf k}})\right] +\ln \left[ 1+\exp (-\beta \widetilde{%
\varepsilon }_{f})\right] \right]  \nonumber \\
&&+\Lambda (R-1)  \label{Eqn.19}
\end{eqnarray}
is the $\Omega ^{u}$ for $\overline{V}_{j,{\bf k},\sigma }=0$. In our case
the unperturbed Hamiltonian for the lattice problem is

{\ 
\begin{eqnarray}
&&H^{u}=\sum_{{\bf k},\sigma }\ \varepsilon _{{\bf k},\sigma }\ c_{{\bf k}%
,\sigma }^{\dagger }c_{{\bf k},\sigma }  \nonumber \\
&&+\sum_{j,\sigma }\widetilde{\varepsilon }_{f}\ f_{{j},\sigma }^{\dagger
}f_{{j},\sigma }+N_{s}\Lambda (R-1)  \nonumber \\
&&+\sum_{j,{\bf k},\sigma }\left( \overline{V}_{j,{\bf k},\sigma
}f_{j,\sigma }^{\dagger }\ c_{{\bf k},\sigma }+\overline{V}_{j,{\bf k}%
,\sigma }^{\ast }c_{{\bf k},\sigma }^{\dagger }\ f_{j,\sigma }\right) ,
\label{Eqn.20}
\end{eqnarray}
and in the impurity case the sum over sites reduces to the impurity site and 
}$N_{s}\Lambda (R-1)\rightarrow \Lambda (R-1)${. This Hamiltonian can be
easily diagonalized, and the corresponding }${\cal H}^{u}$ can be written 
\begin{equation}
{\cal H}^{u}=\sum_{i,\sigma }\omega _{i,\sigma }\ \alpha _{i,\sigma
}^{\dagger }\alpha _{i,\sigma }\ +\Lambda (R-1),  \label{Eqn.21}
\end{equation}
where $\alpha _{i,\sigma }^{\dagger }$ ($\alpha _{i,\sigma }$) are the
creation (destruction) operators of the composite particles of energies $%
\omega _{i,\sigma }$ (there are $N_{s}+1$ states for each spin $\sigma $).
The calculation of 
\begin{equation}
\Omega ^{u}=\frac{-1}{\beta }\sum_{i{\Bbb ,\sigma }}\ln \left[ 1+\exp
(-\beta \ \omega _{i,\sigma })\right] +\Lambda (R-1),  \label{Eqn.22}
\end{equation}
is straightforward, and employing Eqs.~(\ref{Eqn.15},\ref{Eqn.16},\ref
{Eqn.19},\ref{Eqn.22}) we find 
\begin{equation}
\Omega =\overline{\Omega }_{0}+\frac{-1}{\beta }\sum_{i{\Bbb ,\sigma }}\ln %
\left[ 1+\exp (-\beta \ \omega _{i,\sigma })\right] +\Lambda (R-1),
\label{Eqn.23}
\end{equation}
where 
\begin{equation}
\overline{\Omega }_{0}\equiv \Omega _{o}-\Omega _{o}^{u}=-\frac{1}{\beta }%
\ln \left[ \frac{1+2\exp (-\beta \widetilde{\varepsilon }_{f})}{\left(
1+\exp (-\beta \widetilde{\varepsilon }_{f})\right) ^{2}}\right] ,
\label{Eqn.24}
\end{equation}
and the eigenvalues $\omega _{i,\sigma }$ of the{\ \ }${\cal H}^{u}$ are
just given by the poles of the GF in the CHA (Eq.~(\ref{Eqn.7})).

It is interesting that all the correlation effects on the thermodynamic
potential appear in the $\Omega _{o}$, that corresponds to the unperturbed
system. The effect of the perturbation in the present approximation is to
redistribute the energy of the quasi-particles in the same way as in an
uncorrelated system with hybridization constant $\overline{V}_{j,{\bf k}%
,\sigma }=\sqrt{D_{0\sigma }}V_{j,{\bf k},\sigma }$, and one could then
expect Fermi liquid behavior in the CHA (see a formal discussion in the
Appendix).

{\ The parameter $\Lambda $ is obtained minimizing }${\Omega }${\ with
respect to $R$\cite{Hewson93} (at constant $\mu $ as discussed at the
beginning of this Section). To simplify the calculations we shall consider a
conduction band with a constant density of states and width $W=2D$, and an
hybridization constant }$V_{\sigma }({\bf k})=V${. We then obtain\ 
\begin{equation}
\frac{\partial \Omega }{\partial R}=\sum_{i{\Bbb ,\sigma }}n_{F}(\omega
_{i,\sigma })\ \left( \frac{\partial \omega _{i,\sigma }}{\partial R}\right)
+\Lambda =0.  \label{Eqn.25}
\end{equation}
}

{The poles of the impurity GF satisfy: } {\ 
\begin{equation}
\omega _{i,\sigma }=\varepsilon _{f}+\left( \frac{V^{2}D_{\sigma }}{2D}%
\right) \ln \left[ \frac{\omega _{i,\sigma }+D}{\omega _{i,\sigma }-D}\right]
,
\end{equation}
} {\ and calculating the $R$ derivative of $\omega _{i}(\vec{k},\sigma )$ we
obtain } {\ 
\begin{equation}
\Lambda =\frac{-V^{2}}{2D}\sum_{i\sigma }\ln \left( \frac{\omega _{i,\sigma
}+D}{\omega _{i,\sigma }-D}\right) \frac{(\omega _{i,\sigma }^{2}-D^{2})%
\hspace{0.1cm}n_{F}(\omega _{i,\sigma })}{(\omega _{i,\sigma
}^{2}-D^{2}+V^{2}D_{\sigma })}.
\end{equation}
} {\ In the absence of magnetic field, i.e. when the }$\omega _{i,\sigma
}=\omega _{i}$ is independent of $\sigma $, {we can write this equation as: }
{\ 
\begin{eqnarray}
\Lambda &=&\frac{-V^{2}}{D}\int_{-\infty }^{\infty }d\omega \hspace{0.1cm}%
\rho _{f}(\omega )\ln \left( \frac{\omega +D}{\omega -D}\right)  \nonumber \\
&&\times \frac{(\omega ^{2}-D^{2})\hspace{0.1cm}n_{F}(\omega )}{(\omega
^{2}-D^{2}+V^{2}D_{\sigma })},  \label{Lambda}
\end{eqnarray}
} {\ \noindent where $\rho _{f}(\omega )=\sum_{i}\delta ($}$\omega -\omega
_{i})${\ is the density of }$f$ {states, and the chemical potential }$\mu ${%
\ of the electrons was included into the single particle energies (cf. Eqs. (%
\ref{E2.3a},\ref{E2.3b})). } {\ The Kondo temperature \ 
\begin{equation}
T_{K}\simeq D\exp \left( \frac{\widetilde{\varepsilon }_{f}}{\overline{V}^{2}%
}\right) ,
\end{equation}
is defined as the }${T}$ that makes{\ the slave boson parameter $r$ to
vanish, but another useful definition was given by Bernhard and Lacroix \cite
{Lacroix99} by taking $T_{K}$ as the crossover temperature,} determined by
the maximum of the derivative of $\left\langle c_{i\sigma }^{\dagger
}f_{i\sigma }\right\rangle $ with respect to $T$.

\section{Impurity results}

{\ \label{sec5} }

{\ The results presented in this section correspond to a Kondo regime with
the following parameters: $Ef=-0.15$; $W=2.0$, $V=0.3$. }

\begin{figure}[ht]
\centering
\epsfysize=2.3truein \centerline  {\epsfbox{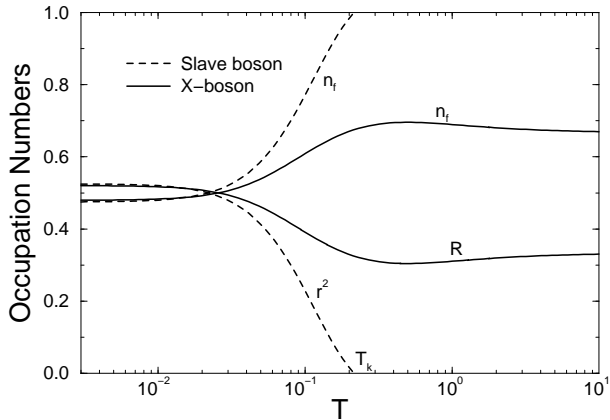}}
\caption{Occupation numbers $n_{f}$, and parameters $r^{2}$ and $R$
as a function of $T$ for both the slave boson and the X-boson methods. The
figures correspond to the following parameters: $E_{f}=-0.15$; $W=2.0$; $%
V=0.3$; $\protect\mu=0.0$.}
\label{figimpu1}
\end{figure}

{\ Figure~\ref{figimpu1} shows the evolution with }$T${\ of the parameter $R$%
, that measures the hole occupation, together with the corresponding }$r^{2}$
{of the usual mean field slave boson, as well as the two occupation number} $%
n_{f}${. The figure shows that }$r^{2}\rightarrow 0$ at a finite $T${\ \
(Kondo temperature $T_{K}$) while }$R$ remains positive{, avoiding the
spurious phase transition. The }$n_{f}\rightarrow 1$ at $T_{K}$ for the
slave boson, while it remains lower than 1 for the X-boson at all $T$.

{At high temperatures the system is well described by localized moments
coupled to the conduction electrons through a Coqblin-Schriefer type
Hamiltonian, and even at high temperatures there should not be a complete
decoupling nor a second order phase transition, as suggested by the mean
field slave-boson theory. \cite{Fulde88} As the GF of the CHA retain the
correlation properties of the system, the X-boson approach shows no
decoupling nor spurious phase transitions, as discussed in Section \ref
{Sec3.3}. One can see in Figure~\ref{figimpu1} that the two approaches give
similar results at low temperatures.}

\begin{figure}[ht]
\centering
\epsfysize=2.2truein \centerline  {\epsfbox{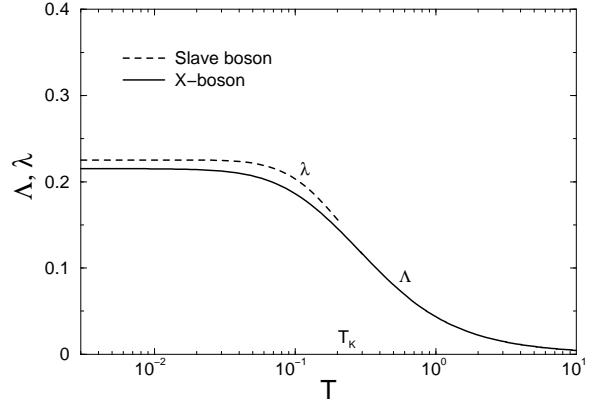}}
\caption{Evolution with temperature $T$ of the renormalizing $%
\Lambda $ and $\protect\lambda$, for the same model parameters used in
Figure~\ref{figimpu1}. }
\label{figimpu2}
\end{figure}

\begin{figure}[ht]
\centering
\epsfysize=2.3truein \centerline  {\epsfbox{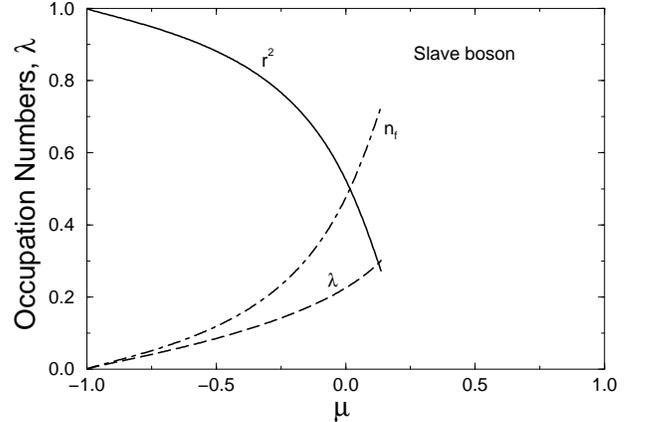}}
\caption{Slave boson: $\protect\lambda$, $r^{2}$, and $n_{f}$ vs. $%
\protect\mu$ for the same model parameters used in Figure~\ref{figimpu1} and 
$T=0.001 $.}
\label{figimpu3}
\end{figure}

\begin{figure}[ht]
\centering
\epsfysize=2.2truein \centerline  {\epsfbox{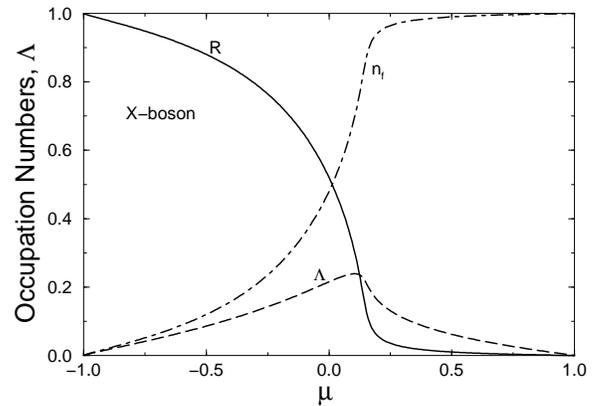}}
\caption{X-boson: $\Lambda$, $R$, and $n_{f}$ vs. $\protect\mu$
for the same model parameters used in Figure~\ref{figimpu3}.}
\label{figimpu4}
\end{figure}

{\ In Figure~\ref{figimpu2} we represent the parameters $\lambda$ and $%
\Lambda$ as a function of temperature. We observe that the slave boson $%
\lambda $ breaks down at the Kondo temperature $T_{K}$ whereas the X-boson $%
\Lambda$ goes continuously to zero. In the high temperature limit the
results obtained with the usual CHA are recovered by the X-boson. }
\ In Figure~\ref{figimpu3} we show the evolution of $\lambda $, $r^{2}$, and 
$n_{f}$, as a function of the chemical potential $\mu $ in the usual slave
boson approach: the formalism breaks down at a value $\mu _{0}$, where $%
n_{f}\rightarrow 1$. In Figure~\ref{figimpu4} we show the evolution of the
parameters $\Lambda $, $R$, and $n_{f}$ as a function of the chemical
potential $\mu $ for the X-boson approach. The results of the two approaches
are similar for $\mu <\mu _{0}$, i.e. before the breakdown of the slave
boson method (compare Fig.~\ref{figimpu3} and Fig.~\ref{figimpu4}). The
parameter $\Lambda $ reaches a maximum in the Kondo region and goes to zero
when $\mu >>E_{f}$, where $n_{f}\rightarrow 1$, and in this limit we recover
the usual CHA.

\begin{figure}[ht]
\centering
\epsfysize=2.5truein \centerline  {\epsfbox{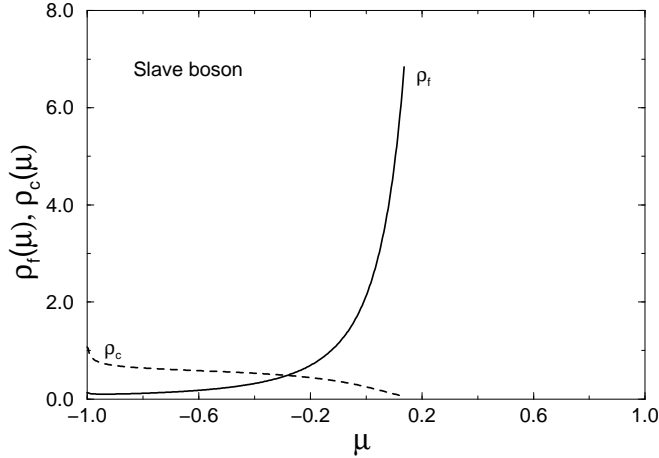}}
\caption{Slave boson: $\protect\rho_{f}(\protect\mu)$, $\protect%
\rho_{c}(\protect\mu)$ vs. $\protect\mu$ for the same model parameters used
in Figure~\ref{figimpu3}.}
\label{figimpu5}
\end{figure}

\begin{figure}[ht]
\centering
\epsfysize=2.5truein \centerline  {\epsfbox{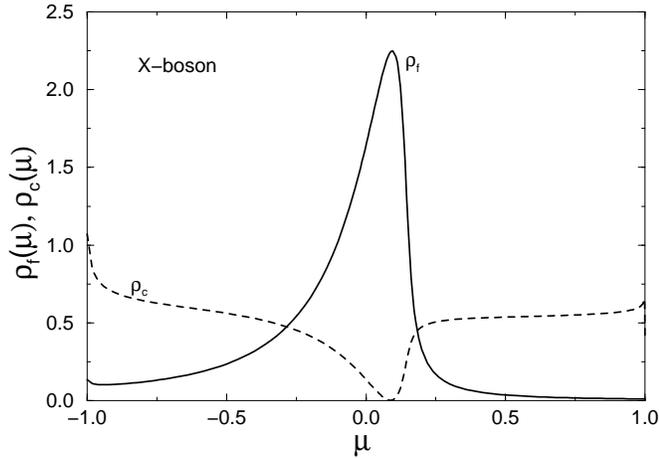}}
\caption{X-boson: $\protect\rho_{f}(\protect\mu)$, $\protect\rho%
_{c}(\protect\mu)$ vs. $\protect\mu$ for the same model parameters used in
Figure~\ref{figimpu3}.}
\label{figimpu6}
\end{figure}

{The values of the f and c electron density of states $\rho _{f}(\mu )$, and 
$\rho _{c}(\mu )$ at }$\omega =\mu $ are shown {as a function of $\mu $ for
the slave boson method in Figure~\ref{figimpu5} and for the X-boson approach
in Figure \ref{figimpu6}. The slave boson plot breaks down at }$\mu =\mu
_{0} ${\ ( where $n_{f}\rightarrow 1)$ while for the X-boson the density of
states $\rho _{f}(\mu )$ is maximum in the Kondo region and goes to zero
when $\mu >>E_{f}$. In both cases one observes the transfer of conduction
electrons to the $f$ band giving rise to the Kondo resonance. }

\vspace{0.6cm}

\vspace{0.6cm}

\begin{figure}[ht]
\centering
\epsfysize=2.5truein \centerline  {\epsfbox{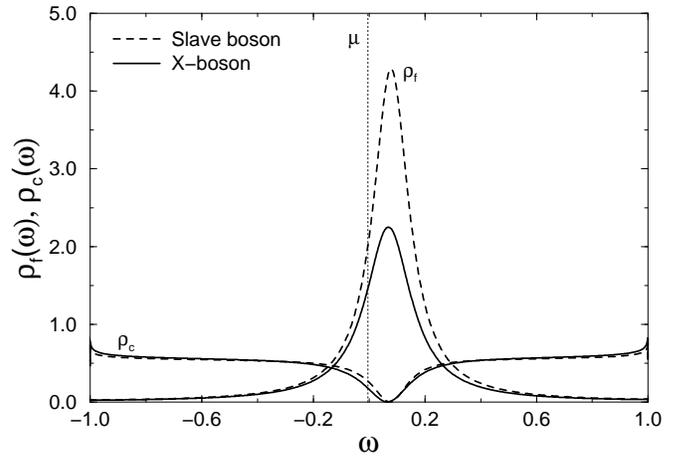}}
\caption{Density of states of $\protect\rho_{f}(\protect\omega)$, $%
\protect\rho_{c}(\protect\omega)$ vs. $\protect\omega$ in a typical Kondo
situation, in the two approaches, for the same model parameters used in
Figure~\ref{figimpu3} and $\protect\mu = 0.0$.}
\label{figimpu7}
\end{figure}

\vspace{0.6cm}
{\ In Figure~\ref{figimpu7} we present the density of states $\rho
_{f}(\omega )$ and $\rho _{c}(\omega )$ vs. energy $\omega $ in a typical
Kondo situation, both for the slave boson and X-boson treatments. The Kondo
resonance appears in the two cases with a similar shape, but it is less
pronounced in the X-boson treatment. }

\section{The lattice problem}

{\ \label{sec6} }

For the lattice we follow the same technique employed with the impurity: {%
the parameter $\Lambda $ is obtained by minimizing the thermodynamic
potential $\Omega =-k_{B}T\ln ({\cal Q})$ with respect to $R$ and ${D}%
_{\sigma }=R+n_{f\sigma }$ is calculated self-consistently. The }$\Omega $ {%
is again obtained from Eq. (\ref{Eqn.15}) }with adequate values of $\Omega
_{0}$ and $<H_{1}(\xi )>_{{\xi }}$:{\ 
\begin{eqnarray}
\Omega _{o} &=&-\frac{2}{\beta }{\sum_{{\bf k}}}\ln \left[ 1+\exp (-\beta
\varepsilon _{{\bf k}})\right] +  \nonumber \\
&&-\frac{N_{s}}{\beta }\ln \left[ 1+2\exp (-\beta \widetilde{\varepsilon }%
_{f})\right] +N_{s}\Lambda (R-1),  \label{Eqn.30}
\end{eqnarray}
\begin{align}
& \left\langle H_{1}\right\rangle _{{\xi }}=\frac{1}{\pi }%
\int\limits_{-\infty }^{\infty }d\omega \ n_{F}(\omega )\   \nonumber \\
\times \sum_{{\bf k,}\sigma }\mathop{\rm Im}& \left\{ \frac{{\xi }\
|V_{\sigma }({\bf k})|^{2}D_{\sigma }}{\left( \omega ^{+}-\varepsilon
_{f\sigma }\right) \left( \omega ^{+}-\varepsilon _{{\bf k}\sigma }\right) -{%
\xi }^{2}\ |V_{\sigma }({\bf k})|^{2}D_{\sigma }}\right\} ,  \label{Eqn.31}
\end{align}
where }$\omega ^{+}=\omega +i0$. As in the impurity case, this expression
coincides with that obtained for an uncorrelated Hamiltonian, which in this
case is Eq. (\ref{Eqn.20}) with a scaled hybridization $\overline{V}_{\sigma
}({\bf k})=\sqrt{D_{0\sigma }}V_{\sigma }({\bf k})$. The corresponding
eigenvalues can be calculated analytically in this case, because for each
spin component $\sigma $ the Hamiltonian is reduced into $N_{s}$ matrices $%
2\times 2$. They are the poles of the lattice GFs, replacing the $N_{s}+1$
eigenvalues $\omega _{i,\sigma }$ \ of the impurity, and we denote them with 
\begin{eqnarray}
&&\omega _{{\bf k},\sigma }(\pm )=\frac{1}{2}\left( \varepsilon _{{\bf k}%
,\sigma }+\widetilde{\varepsilon }_{f}\right)  \nonumber \\
&&\pm \frac{1}{2}\sqrt{\left( \varepsilon _{{\bf k},\sigma }-\widetilde{%
\varepsilon }_{f}\right) ^{2}+4\left| V_{\sigma }({\bf k})\right|
^{2}D_{\sigma }}.  \label{Eqn.32}
\end{eqnarray}
Following the same arguments employed for the impurity we obtain 
\begin{eqnarray}
\Omega &=&\overline{\Omega }_{0}+\frac{-1}{\beta }\sum_{{\bf k},\sigma ,\ell
=\pm }\ln \left[ 1+\exp (-\beta \ \omega _{{\bf k},\sigma }(\ell )\right] 
\nonumber \\
&&+N_{s}\ \Lambda (R-1),  \label{Eqn.33}
\end{eqnarray}
where 
\begin{equation}
\overline{\Omega }_{0}=-\frac{N_{s}}{\beta }\ln \left[ \frac{1+2\exp (-\beta 
\widetilde{\varepsilon }_{f})}{(1+\exp (-\beta \widetilde{\varepsilon }%
_{f}))^{2}}\right] .  \label{Eqn.34}
\end{equation}
The same result was obtained by direct analytical integration, thus
confirming the arguments employed in the derivation of Eqs. (\ref{Eqn.23},%
\ref{Eqn.33}). As in the impurity case, all the correlation effects on the
thermodynamic potential appear in the $\Omega _{o}$, and one expects again a
Fermi liquid behavior of the quasiparticles in the X-boson approximation
(cf. the Appendix).

As in the impurity case we obtain an equation for $\Lambda $ by minimizing $%
\Omega $ with respect of $R$:\ {\ } {\ 
\begin{eqnarray}
\Lambda &=&\frac{1}{N_{s}}\sum_{{\bf k},\sigma }\left| V_{\sigma }({\bf k}%
)\right| ^{2}  \nonumber \\
&&\times \frac{n_{F}\left( \omega _{{\bf k},\sigma }(+)\right) -n_{F}\left(
\omega _{{\bf k},\sigma }(-)\right) }{\sqrt{\left( \varepsilon _{{\bf k}%
,\sigma }-\widetilde{\varepsilon }_{f,\sigma }\right) ^{2}+4\left| V_{\sigma
}({\bf k})\right| ^{2}D_{\sigma }}},  \label{Eqn.35a}
\end{eqnarray}
where $n_{F}\left( x\right) $ is the Fermi-Dirac distribution. Employing }$%
\left| V_{\sigma }({\bf k})\right| =V$ as well as{\ a conduction band of
constant density of states and width $W=2D$, we find } {\ 
\begin{equation}
\Lambda =\frac{V^{2}}{D}\int_{-D}^{D}d\varepsilon _{{\bf k}}\frac{%
n_{F}\left( \omega _{{\bf k}}(+)\right) -n_{F}\left( \omega _{{\bf k}%
}(-)\right) }{\sqrt{\left( \varepsilon _{{\bf k}}-\widetilde{\varepsilon }%
_{f}\right) ^{2}+4V^{2}D_{\sigma }}}.  \label{Eqn.36}
\end{equation}
The quasiparticle properties of a heavy fermion system can be described by
an effective Hamiltonian, characterized by two hybridized bands coupled by
an effective hybridization matrix element $\overline{V}$.\cite{Fulde88} In
the slave boson method 
\begin{equation}
\overline{V}_{S-b}^{2}=(1-n_{f})V^{2},  \label{Eq.13}
\end{equation}
} {\ while in the present approach we have 
\begin{equation}
\overline{V}_{X-b}^{2}=D_{\sigma }V^{2}=(1-n_{f}/2)V^{2}.  \label{Eq.13a}
\end{equation}
} {\ \noindent The hybridization reduction factor is related to the
effective probability that a }$c$ {electron jumps into an }$f$ state{, and
when $U\rightarrow \infty $ this transition can only take place if the $f$
level of the final site is empty. Rice and Ueda\cite{Rice86} argued that the
effective hybridization should be } {\ 
\begin{equation}
{\overline{V}}_{RU}^{2}=\frac{1-n_{f}}{1-n_{f}/2}V^{2},  \label{Eq.14}
\end{equation}
and variational calculations were performed by them\cite{Rice86} and by
Fazekas.\cite{Fazekas87,Fazekasbook} These variational calculations present
the same spurious phase transitions shown by the slave-boson method, because
in all these treatments the reduction factor goes to zero at a critical
temperature. On the other hand, the reduction factor does not vanish in the
whole range of temperatures for the X-boson approach, and the $f$ and $c$
electrons do not decouple. All these approaches produce similar results at
low temperatures. }

\section{Lattice results}

{\ \label{sec7} }

{We have employed for the lattice the same system parameters used to discuss
the impurity, and a similar type of behavior is observed. The evolution of
the X-boson parameters $R$, and }$n_{f}$ are shown i{n Figure \ref{figrede1}
as a function of $T$, and the $f$ and $c$ electrons do not decouple in this
treatment because the }hole occupation{\ $R>0$ in the whole range of
temperatures, while the corresponding parameter $r^{2}$ of the usual slave
boson treatment vanishes at a critical temperature (Kondo temperature $T_{K}$%
).

\begin{figure}[ht]
\centering
\epsfysize=2.2truein \centerline  {\epsfbox{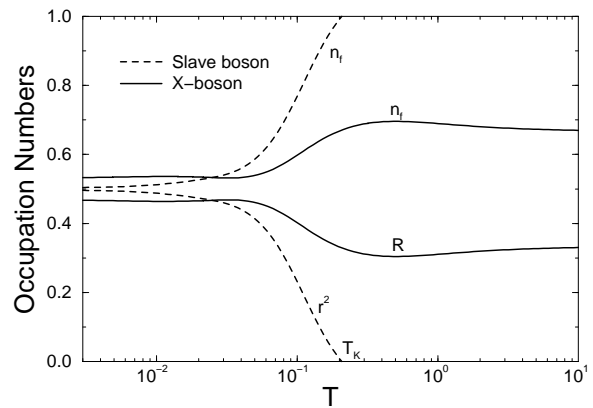}}
\caption{Kondo case: Occupation numbers $n_{f}$, and parameters $%
r^{2}$ and $R$ vs.  $T$ for both the slave boson and the X-boson
methods, for the same model parameters used in Figure~\ref{figimpu1}.}
\label{figrede1}
\end{figure}

\begin{figure}[ht]
\centering
\epsfysize=2.2truein \centerline  {\epsfbox{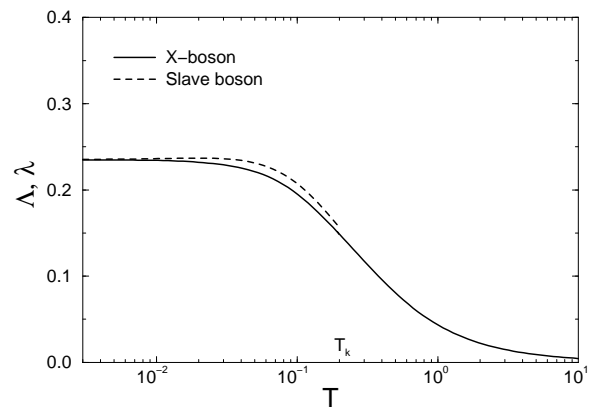}}
\caption{Kondo case: $\Lambda$, and $\protect\lambda$ vs. $T$ for
the same model parameters used in Figure~\ref{figrede1}.}
\label{figrede2}
\end{figure}

The two approaches yield similar results at low temperatures. \ The
dependence with temperature of $\lambda $ (slave boson) and $\Lambda $
(X-boson) is presented in Figure \ref{figrede2}, and }t{he slave boson $%
\lambda $ breaks down at the Kondo temperature $T_{K}$ whereas the X-boson $%
\Lambda $ goes continuously to zero, recovering the CHA behavior at high
temperatures.}

\begin{figure}[ht]
\centering
\epsfysize=2.4truein \centerline  {\epsfbox{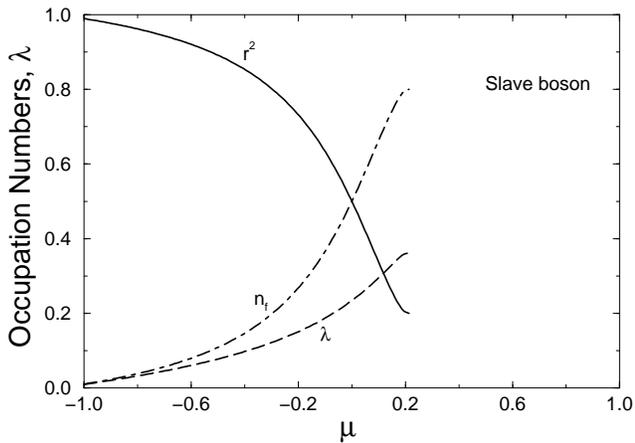}}
\caption{Slave boson: Occupation numbers, $\protect\lambda$, $%
r^{2}$, and $n_{f}$ vs. $\protect\mu$ for the same model parameters used in
Figure~\ref{figimpu3}.}
\label{figrede3}
\end{figure}

\begin{figure}[ht]
\centering
\epsfysize=2.4truein \centerline  {\epsfbox{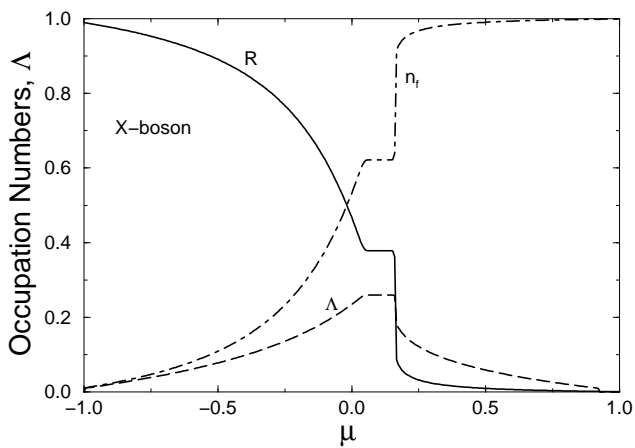}}
\caption{X-boson: Occupation numbers, $\Lambda$, $R$, and $n_{f}$
vs. $\protect\mu$ for the same model parameters used in Figure~\ref{figimpu3}.}
\label{figrede4}
\end{figure}

T{he evolution of }$n_{f}$,{\ $\lambda $ and }$r^{2}$, with{\ the chemical
potential $\mu $, is presented in Figure \ref{figrede3} for the slave boson
treatment. In this case we do not recover the three characteristics regimes
of the PAM: Kondo, intermediate valence and magnetic, because the formalism
breaks down when $n_{f}\rightarrow 1$. The corresponding quantities in the
X-boson approach: }$n_{f}$,{\ }$\Lambda ${ and }$R$,{\ are also given as a
function of $\mu $ in Figure \ref{figrede4}, and the three characteristic
regimes of the PAM are clearly shown. There is a plateau in this figure when
the chemical potential crosses the hybridization gap, and the parameter $%
\Lambda $ is maximum in the Kondo region, and goes smoothly to zero together
with }$R$ when $\mu $ increases, while{\ $n_{f}\rightarrow 1.$ The CHA is
recovered in this region by the X-boson treatment.}

\begin{figure}[ht]
\centering
\epsfysize=2.6truein \centerline  {\epsfbox{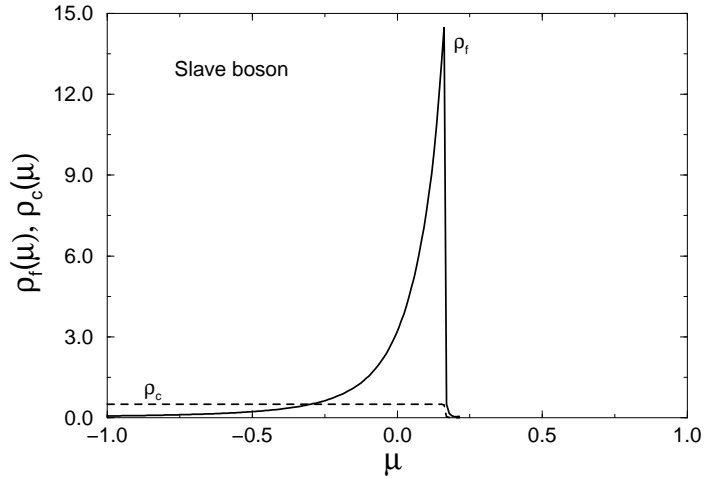}}
\caption{Slave boson: $\protect\rho_{f}(\protect\mu)$, $\protect%
\rho_{c}(\protect\mu)$ vs. $\protect\mu$ for the same model parameters used
in Figure~\ref{figimpu3}.}
\label{figrede5}
\end{figure}

\begin{figure}[ht]
\centering
\epsfysize=2.6truein \centerline  {\epsfbox{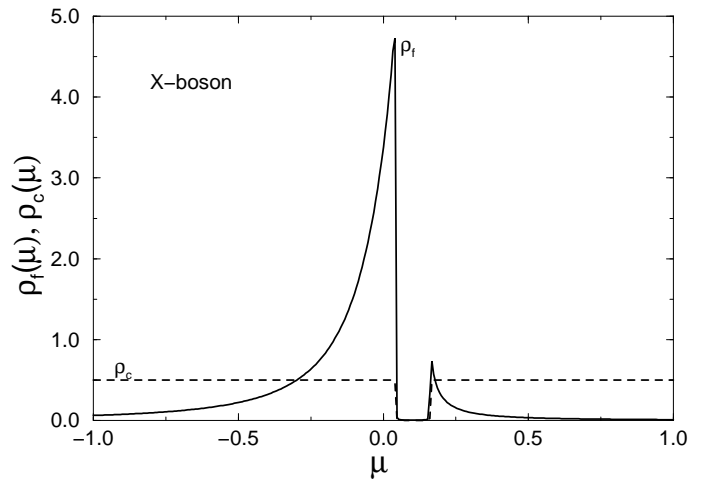}}
\caption{X-boson: $\protect\rho_{f}(\protect\mu)$, $\protect\rho%
_{c}(\protect\mu)$ vs. $\protect\mu$ for the same model parameters used in
Figure~\ref{figimpu3}.}
\label{figrede6}
\end{figure}

The value of the density of states $\rho _{f}(\mu )$, $\rho _{c}(\mu )$ as
function of $\mu $ is presented in Figure~\ref{figrede5} for the slave boson
treatment: the approach breaks down in the Kondo region when $%
n_{f}\rightarrow 1$ and the same quantities are plotted {in Figure~\ref
{figrede6}} for the X-boson approach.

\vspace{1.0cm}
\vspace{1.0cm}

\begin{figure}[ht]
\centering
\epsfysize=2.5truein \centerline  {\epsfbox{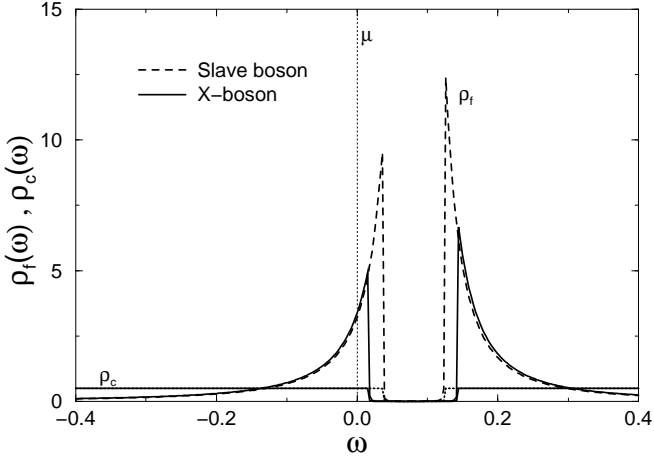}}
\caption{Density of states in a typical Kondo situation in the
two approaches: $\protect\rho(\protect\omega)$ vs $\protect\omega$ for the
same model parameters used in Figure~\ref{figimpu3} and $T=0.001$.}
\label{figrede7}
\end{figure}

\vspace{1.0cm}
{\ The density of states $\rho _{f}(\omega )$, $\rho _{c}(\omega )$ vs $%
\omega $ are shown in Figure \ref{figrede7}\ for both the slave boson and
X-boson approaches in a typical Kondo situation. The density of $f$ states
at $\mu $ is practically the same in the two cases. }

\section{Specific Heat}

{\ \label{Sec7} }

{\ In this section we apply the theory developed in this work to calculate
the specific heat of the Kondo insulators, but for simplicity, we shall not
try to fit the experimental results of some particular compound. This class
of materials has been extensively studied in the last decade since its
characterization by G. Aeppli and Z. Fisk \cite{Aeppli92} as Kondo
insulators. Some of the compounds that we can include in this family are: $%
{\rm FeSi}$; ${\rm Ce}_{3}{\rm Bi}_{4}{\rm Pt}_{3}$; ${\rm SmB}_{6},$ and $%
{\rm YbB}_{12}$. The slave-boson was applied to Kondo insulators initially
by P. Riseborough \cite{Riseborough92} and by Carlos Sanchez-Castro et al,
\cite{Bedell93} and we have recently studied the Kondo insulator ${\rm FeSi}$
using the atomic model. \cite{FoglioFeSi,FeSi2000} We have been able to
adjust simultaneously the static conductivity, the resistivity and the
dynamical conductivity to the experimental results, and we obtained a fair
agreement. In this section, we present the X-boson formalism as an
alternative that gives results close to those obtained by the slave boson
method in its region of validity, while it makes possible to extend the
results to the whole range of temperatures, without presenting the spurious
phase transitions appearing in the usual slave-boson method. }

{\ To calculate the specific heat employing }$\Omega $ we first show by
standard thermodynamic techniques that{\ 
\begin{equation}
S=-\left( \frac{\partial F}{\partial T}\right) _{N,V}=-\left( \frac{\partial
\Omega }{\partial T}\right) _{\mu ,V},  \label{Eqn.40}
\end{equation}
} and then that 
\begin{eqnarray}
C_{v} &=&T\left( \frac{\partial S}{\partial T}\right) _{N,V}=-T\ \left( 
\frac{\partial ^{2}\Omega }{\ \partial T^{2}}\right) _{\mu ,V}  \nonumber \\
&&+T\ \left( \frac{\partial \mu }{\partial T}\right) _{N,V}\ \left( \frac{%
\partial N}{\partial T}\right) _{\mu ,V}.  \label{Eqn.41}
\end{eqnarray}
{\ \noindent Assuming a conduction band with constant density of states and
width $W=2D$ and the absence of magnetic field, we find from Eqs. (\ref
{Eqn.33},\ref{Eqn.34}) that} 
\begin{eqnarray}
&&-T\ \left( \frac{\partial ^{2}\Omega }{\ \partial T^{2}}\right) _{\mu
,V}=-T\ \left( \frac{\partial ^{2}\overline{\Omega }_{0}}{\ \partial T^{2}}%
\right) _{\mu ,V}  \nonumber \\
&&+\frac{k_{B}\hspace{0.1cm}\beta ^{2}}{D}\sum_{\ell =\pm
}^{2}\int_{-D}^{D}dx\hspace{0.1cm}\omega _{\ell }^{2}(x)\hspace{0.1cm}%
n_{F}\left( \omega _{\ell }(x)\right) \left[ 1-n_{F}\left( \omega _{\ell
}(x)\right) \right]  \nonumber \\
&&-T\ N_{s}\left( \frac{\partial ^{2}(\Lambda (R-1))}{\partial T^{2}}\right)
_{\mu ,V},  \label{Eqn.43}
\end{eqnarray}
where 
\begin{equation}
\omega _{\pm }(x)=\frac{1}{2}\left( x+\widetilde{\varepsilon }_{f}\right)
\pm \frac{1}{2}\sqrt{\left( x-\widetilde{\varepsilon }_{f}\right)
^{2}+4\left| V\right| ^{2}D_{\sigma }}  \label{Eqn.44}
\end{equation}
and 
\begin{eqnarray}
&&-T\ \left( \frac{\partial ^{2}\overline{\Omega }_{0}}{\ \partial T^{2}}%
\right) _{\mu ,V}=-2N_{s}\ k_{B}\hspace{0.1cm}\beta ^{2}\widetilde{%
\varepsilon }_{f}^{2}\hspace{0.1cm}\exp (\beta \widetilde{\varepsilon }_{f})
\nonumber \\
&&\qquad \times \frac{[3+2\exp (\beta \widetilde{\varepsilon }_{f})]}{\left[
\exp (\beta \widetilde{\varepsilon }_{f})+2\right] ^{2}\left[ \exp (\beta 
\widetilde{\varepsilon }_{f})+1\right] ^{2}}.  \label{Eqn.45}
\end{eqnarray}

\begin{figure}[ht]
\centering
\epsfysize=2.5truein \centerline  {\epsfbox{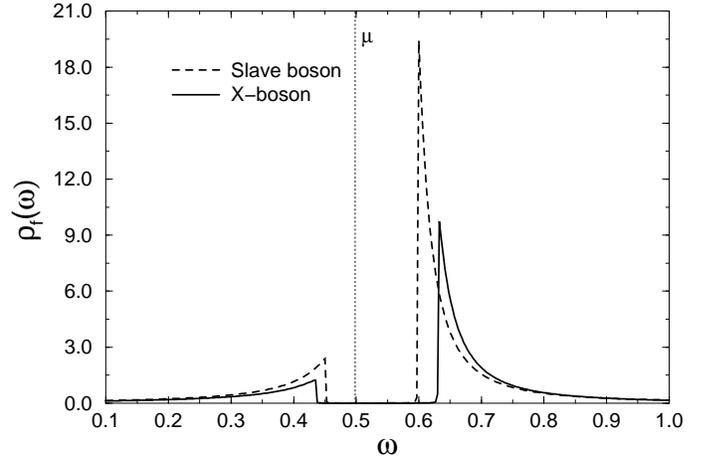}}
\caption{Density of states in a typical Kondo insulator situation
in the two approaches: $\protect\rho_{f}(\protect\omega)$ vs. $\protect\omega
$ with the following parameters: $E_{f}=0.3$; $W=2.0$; $V=0.35$; $\protect\mu%
=0.5$; $T=0.001$.}
\label{figinsu1}
\end{figure}

{\ We compare the specific heat obtained by the slave boson vs. X-boson
methods employing in the two cases the same parameters, corresponding to a
typical Kondo insulator situation with the chemical potential inside the
gap. In Figure~\ref{figinsu1} we present the corresponding density of states 
$\rho _{f}(\omega )$ vs. $\omega $, for the following parameters: $E_{f}=0.3$%
; $V=0.35$; $W=2.0$; $T=0.001$ and $\mu =0.5$.}

\begin{figure}[ht]
\centering
\epsfysize=2.5truein \centerline  {\epsfbox{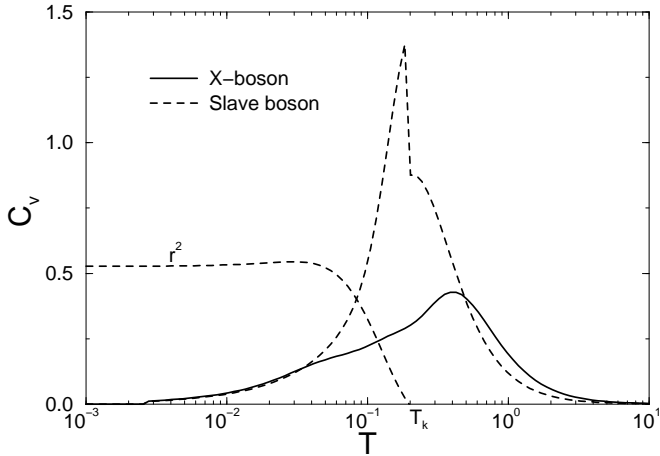}}
\caption{Kondo insulator: Specific heat of a typical Kondo
insulator for the same model parameters used in Figure~\ref{figinsu1} but as
function of temperature.}
\label{figinsu2}
\end{figure}

In Figure~\ref{figinsu2} we present $C_{v}$ vs. $T$, employing the same
parameters of Figure~\ref{figinsu1}. We have calculated 
\mbox{$T\ \left( \partial \mu /\partial T\right) _{N,V}\ \left(
\partial N/\partial T\right) _{\mu ,V}$} numerically, and its contribution
to $C_{V}$ is negligible for these parameters. The calculation above the
Kondo temperature $T_{K}$ in the slave boson case was performed for the
phase of uncoupled $f$ and $c$ electrons, and the specific heat presents a
discontinuity at the transition $T_{K}$, but the curve obtained by the
X-boson method is continuous.

\section{Conclusions}

{\ \label{Sec8} The slave-boson formalism\cite{Coleman84} has been
extensively used in the mean-field approximation\cite{Read87,Hewson93} to
study strongly correlated electron systems, because it captures the
essential physics of the Kondo effect at low temperatures and its
implementation in the Kondo limit is very simple. One drawback of this
method is that above a temperature }$T_{K}$ (cf. Section~ \ref{Sec4}) or {%
when $\mu >>E_{f}$, }it develops a spurious second order phase transition
into a phase with decoupled conduction and localized electrons. In the
present work we present an approach that was inspired in some of the
features of the slave-boson method in the mean-field approximation and gives
similar result at low temperatures, but it does not have the spurious phase
transitions of that method at higher temperatures or chemical potentials,
where it behaves like the CHA solution. We have named ``$X$-boson'' this new
technique because it keeps the $X$ operators\cite{Hubard4} even at the final
stages of the calculation, while two hybridized but uncorrelated bands
appear in the slave-boson method at that stage. What we employ instead is an
approximation to the exact cumulant expansion of the PAM in the limit of $%
U\rightarrow \infty $, which is called{\ CHA (chain approximation).\cite
{FFM,Infinite} The CHA corresponds to the most general set of diagrams with
only second order cumulants, and has interesting properties like being $\Phi
-$derivable.\cite{ChainPhi} We have developed the X-boson method both for a
single local impurity and for a lattice. The correlation appears in this
approximation only through the presence of a constant }$D_{0\sigma }$ at
critical places of the corresponding GF (cf. Eqs. (\ref{Eqn.7}\ -\ \ref
{Eqn.9b},\ \ref{Eqn.10a}\ -\ \ref{Eqn.13})), but it has profound effect on
the system description, forcing the local electron occupation to be $n_{f}<1$%
, and eliminating the spurious phase transitions of the slave-boson method.

{From the computational point of view the X-boson method is equivalent to
the slave-boson technique, and the possibility of using it for all
temperatures or chemical potentials }$\mu $ makes it useful as {the starting
point to study several interesting problems: competition between Kondo
effect and RKKY interaction in heavy fermion systems,\cite
{Iglesias97,Tatiana2000,Castro2000} impurity bands in Kondo insulators,\cite
{Schlottmann92} non Fermi-liquid behavior in disordered systems,\cite
{Miranda96} and Kondo effect in quantum dots.\cite{Craco99,Enrique99} To give 
an example we have considered a Kondo insulator
and calculated its specific heat: the slave boson method shows a
discontinuity at }$T_{K}$, while {it is continuous in the X-boson method.
Although it is }$C_{p}$ that is usually measured, the difference $%
C_{p}-C_{v} $ is usually small in liquids and solids, and shows a dependence
with $T\,\ $similar to that obtained with {the X-boson method (see for
example, the $C_{p}$ of $FeSi$ measured by Jacarino and coworkers\cite
{Jacarino67}).}

{\ In conclusion, we have presented an approach that gives essentially the
same results as Coleman's slave-boson treatment at low temperatures, but
without the spurious phases transitions at intermediate temperatures }$T${\
or large values of $\mu $. The X-boson results approach those of the
cumulant expansion in the chain approximation for large values of }$T$ or $%
\mu $.{\ }

{\ {\bf Acknowledgments.} We acknowledge Profs. E. Miranda (UNICAMP) 
and E. V. Anda (PUC-RJ)
for helpful discussions and the financial support from S\~{a}o Paulo State
Research Foundation (FAPESP) and National Research Council (CNPq). }

{\ \appendix{\ } }

\section{Fermi liquid behavior of the X-boson Method}

{\label{Ap1}We show here that the PAM has Fermi liquid properties in the X
boson description, and to discuss the problem we shall first consider the ${f%
}$ and $c$ electrons self-energies $\Sigma _{f}({\bf k},z)$ and $\Sigma _{c}(%
{\bf k},z)$, that are related to the exact GF by the equations {\ 
\begin{equation}
G_{{\bf k}{\sigma }}^{ff}(z)=\frac{-1}{(z-\widetilde{\varepsilon }_{f,\sigma
}-\Sigma _{f}({\bf k},z))},  \label{Eq.20}
\end{equation}
} {\ 
\begin{equation}
G_{{\bf k}{\sigma }}^{cc}(z)=\frac{-1}{(z-\varepsilon _{{\bf k},\sigma
}-\Sigma _{c}({\bf k},z))},  \label{Eq.21}
\end{equation}
where we employ the }}$z$ with the $\mu $ {already subtracted} (cf. Eqs.(\ref
{E2.3a},\ref{E2.3b})). {{Employing an effective cumulant\cite
{FoglioFeSi,FeSi2000} }$M_{2,{\bf k}\sigma }^{eff}(z)$ {it is possible to
write the exact GF} 
\begin{equation}
G_{{\bf k}{\sigma }}^{ff}(z)=\frac{{M_{2,{\bf k}\sigma }^{eff}(z)}}{1-|V(%
\vec{k})|^{2}G_{c,{\bf k}\sigma }^{o}(z)\ {M_{2,{\bf k}\sigma }^{eff}(z)}}
\label{Eq.25}
\end{equation}
and 
\begin{eqnarray}
G_{{\bf k}{\sigma }}^{cc}(z) &=&\frac{G_{c,{\bf k}\sigma }^{o}(z)}{1-|V({\bf %
k})|^{2}G_{c,{\bf k}\sigma }^{o}(z)\ {M_{2,{\bf k}\sigma }^{eff}(z)}} 
\nonumber \\
&=&\frac{-1}{z-\varepsilon _{{\bf k}\sigma }+|V({\bf k})|^{2}\ {M_{2,{\bf k}%
\sigma }^{eff}(z)}},  \label{Eq.26}
\end{eqnarray}
where 
\begin{equation}
G_{c,{\bf k}\sigma }^{o}(z)=\frac{-1}{(z-\varepsilon _{{\bf k}\sigma })}
\label{Eq.210}
\end{equation}
is the unperturbed GF of the $c$ electrons. The exact self energies are then{%
\ 
\begin{equation}
\Sigma _{f}({\bf k},z)=z-\widetilde{\varepsilon }_{f,\sigma }+[{M_{2,{\bf k}%
\sigma }^{eff}(z)}]^{-1}-|V({\bf k})|^{2}G_{c,{\bf k}\sigma }^{o}(z)
\label{Eq.28}
\end{equation}
and 
\begin{equation}
\Sigma _{c}({\bf k},z)=-|V({\bf k})|^{2}\ {M_{2,{\bf k}\sigma }^{eff}(z)},
\label{Eq.29}
\end{equation}
} } and to recover the CHA self energies it is sufficient to replace{\cite
{Infinite} the effective cumulant $M_{2,{\bf k}\sigma }^{eff}(z)$} by the
contribution of its simplest diagram, namely the unperturbed GF of the $f$
electrons: 
\begin{equation}
{M_{2,{\bf k}\sigma }^{eff}(z)}\rightarrow {G_{f,0\sigma
}^{o}(z)=-D_{0\sigma }/{(z-\varepsilon _{\sigma }).}}  \label{Eq.30}
\end{equation}

The essential property of {a Fermi liquid is that at low temperatures it has
a behavior close to that of a system of free Fermions. It is described by a
system of quasi-particles that replace the elementary excitations of the
free system, but they have a finite life-time caused by the interactions
that are absent in the free system. Instead of the elementary particle
energies we have to analyze} the poles of the relevant GF, given in our
problem by the solutions $z_{{\bf k\sigma }}$ of the equation 
\begin{equation}
z-\varepsilon _{{\bf k}\sigma }+|V({\bf k})|^{2}\ {M_{2,{\bf k}\sigma
}^{eff}(z)=0.}  \label{Eq.31}
\end{equation}
The $%
%TCIMACRO{\func{Re}}%
%BeginExpansion
\mathop{\rm Re}%
%EndExpansion
\left[ z_{{\bf k\sigma }}\right] $ corresponds to the energies of the
elementary excitations, while their imaginary parts give their decay
properties, and the Fermi surface is given by the set of ${\bf k}_{F}$ such
that $%
%TCIMACRO{\func{Re}}%
%BeginExpansion
\mathop{\rm Re}%
%EndExpansion
\left[ z_{{\bf k\sigma }}\right] =0$, because we already have subtracted the 
$\mu $ from the $z$. The ${M_{2,{\bf k}\sigma }^{eff}(z)}$ \ in the
numerator of Eq. (\ref{Eq.25}) does not introduce new poles, because $G_{%
{\bf k}{\sigma }}^{ff}(z)$ is finite at those poles.

>From Eq. (\ref{Eq.31}) we obtain 
\[
{M_{2,{\bf k}\sigma }^{eff}(z_{{\bf k\sigma }})=-}\frac{z_{{\bf k\sigma }%
}-\varepsilon _{{\bf k}\sigma }}{|V({\bf k})|^{2}}, 
\]
and replacing in Eq. (\ref{Eq.28}) we find 
\begin{equation}
\Sigma _{f}({\bf k},z_{{\bf k\sigma }})=z_{{\bf k\sigma }}-\widetilde{%
\varepsilon }_{f,\sigma },  \label{Eq.32a}
\end{equation}
while from Eqs. (\ref{Eq.30},\ref{Eq.31}) we obtain 
\begin{equation}
\Sigma _{c}({\bf k},z_{{\bf k\sigma }})=z_{{\bf k\sigma }}-\varepsilon _{%
{\bf k}\sigma }.  \label{Eq.32b}
\end{equation}
It then follows that 
\begin{equation}
%TCIMACRO{\func{Im}}%
%BeginExpansion
\mathop{\rm Im}%
%EndExpansion
\left[ \ \Sigma _{c}({\bf k},z_{{\bf k\sigma }})\right] =%
%TCIMACRO{\func{Im}}%
%BeginExpansion
\mathop{\rm Im}%
%EndExpansion
\left[ \ \Sigma _{f}({\bf k},z_{{\bf k\sigma }})\right] =%
%TCIMACRO{\func{Im}}%
%BeginExpansion
\mathop{\rm Im}%
%EndExpansion
\left[ z_{{\bf k\sigma }}\right] .  \label{Eq.33}
\end{equation}

{An essential property of the Fermi liquid is} that the quasi-particles on
the Fermi surface{\ have an infinite lifetime} when $T\rightarrow 0$, and
this property follows from Eq. (\ref{Eq.33}) for both the $f$ and $c$
electrons when the poles on the Fermi surface are real, because the
corresponding self-energies are then also real.

>From Eq. (\ref{Eqn.32}) we conclude that the $GF$ poles of the PAM in the
CHA are all real, and the same property holds for the X-boson solution.
These two approximations therefore describe the PAM as a Fermi liquid,
because they satisfy the condition above. Because the infinite lifetime is
valid for all ${\bf k}$ and for all $T$, the quasi-particles in these two
approximations behave more like elementary excitations than like the usual
quasi-particles, that have a finite lifetime outside the Fermi surface, even
at $T\rightarrow 0$.

\end{document}